# Excitation and coherent control of magnetization dynamics in magnetic tunnel junctions using acoustic pulses

H. F. Yang,[1] F. Garcia-Sanchez,[1,2] X. K. Hu,[1] S. Sievers,[1] T. Böhnert,[3] J. D. Costa,[3*]
M. Tarequzzaman,[3] R. Ferreira,[3] M. Bieler,[1+] and H. W. Schumacher[1]

*[1]Physikalisch-Technische Bundesanstalt, Bundesallee 100,*
*D-38116 Braunschweig, Germany*

*[2]Istituto Nazionale di Ricerca Metrologica, Strada delle Cacce 91,*
*10135 Torino, Italy*

*[3]International Iberian Nanotechnology Laboratory, Av. Mestre José Veiga,*
*4715-330 Braga, Portugal*

We experimentally study magnetization dynamics in magnetic tunnel junctions driven by femtosecond-laser-induced surface acoustic waves. The acoustic pulses induce a magnetization precession in the free layer of the magnetic tunnel junction through magnetoelastic coupling. The frequency and amplitude of the precession shows a pronounced dependence on the applied magnetic field and the laser excitation position. Comparing the acoustic-wave-induced precession frequencies with precession induced by charge currents and with micromagnetic simulations we identify spatially non-uniform magnetization modes localized close the edge regions as being responsible for the optically induced magnetization dynamics. The experimental scheme even allows us to coherently control the magnetization precession using two acoustic pulses. This might prove important for future applications requiring ultrafast spin manipulation. Additionally, our results directly pinpoint the importance of acoustic pulses since they could be relevant when investigating optically-induced temperature effects in magnetic structures.

---

* Present address: IMEC, 3000 Leuven, Belgium
+ Electronic address: mark.bieler@ptb.de

Manipulating the spin through external stimulus is a key issue in the field of spintronic with the aim to boost logic and memory applications. Such a manipulation can be achieved by different physical effects, employing photons, electrons, heat flux, THz radiation, as well as phonons.[1–8] In particular, the interaction of acoustic pulses with spin structures provides interesting prospects. This is because acoustic pulses can be easily generated on picosecond time scales and the magnetoelastic effect (the change of magnetic properties due to elastic deformation) governing the interaction may lead to significant magnetization changes. Recently, the influence of surface acoustic waves (SAWs) on certain nanoelements and magnetic bubbles was studied.[6,9] Additionally, laser-induced acoustic pulses were used to excite magnetization dynamics in ferromagnetic layers.[10,11] It has been found that the acoustic-pulse-induced precession can be enhanced when being resonantly driven,[12–14] yet, an identification of the precession modes is difficult, even in cases without acoustic perturbation.[15–17] Based upon the previous studies on magnetoelastic effects,[6,9–16] the next logical step would be to extend these studies to magnetic devices having important industrial relevance. Magnetic tunnel junctions (MTJs) are certainly among these devices as they are used in data storage, magnetic sensor and other spintronic applications. So far manipulation of the magnetization in MTJs has been realized by charge currents or heat currents through spin transfer torque (STT).[18,19] Yet, no experiment on SAW-induced magnetization dynamics in MTJs has been reported.

Here, we study the excitation of magnetization dynamics in MTJs by femtosecond-laser-induced SAW taking advantage of the magnetoelastic coupling. Due to the dependence of the tunnel resistance on magnetization orientation of the free magnetic layer in the MTJs, our technique can directly measure small spin precession angles. Using a time resolved detection technique we are able to pinpoint acoustic pulses as being responsible for the spin manipulation and exclude other effects resulting from laser pulse excitation. Moreover, the magnetization mode driven by the SAWs in the free layer of the MTJ can be determined by comparison to magnetization modes triggered by charge current pulses and micromagnetic simulations. So far, the identification of the exact magnetization mode in previous magnetoelastic experiments has not been accomplished. Taking advantage of the coherent nature of SAWs, we also show that our scheme allows coherent control of the magnetization. Using two separate acoustic pulses we can either enhance or switch off the precession in the MTJ.

The experiments were carried out on a rectangular MTJ nanopillar stack with lateral dimensions of 100 nm × 550 nm. The stack was deposited on a Si wafer and consists (from bottom to top) of an antiferromagnet (20 nm IrMn), a synthetic antiferromagnet (2 nm $Co_{70}Fe_{30}$, 0.85 nm Ru, 2.6 nm $Co_{40}Fe_{40}B_{20}$), the tunnel junction (~0.8 nm MgO, corresponding to a resistance-area product of 1.8 $\Omega\mu m^2$), and the free layer (2.6 nm $Co_{40}Fe_{40}B_{20}$). Between the Si wafer and the MTJ stack a 100 nm thick $Al_2O_3$ layer and a CuN layer serve as isolation layer and bottom contact, respectively. On top of the MTJ a 30 nm thick $(Ti_{10}W_{90})_{100-x}N_x$ layer and a 300 nm thick Al layer were deposited and patterned, serving as top contact and transducer to convert ultrafast laser pulses into acoustic phonon pulses, see Fig. 1(a). The tunnel magnetoresistance of the MTJ is approximately 100% and its magnetic anisotropy field is 100 Oe as being determined from a measured Stoner-Wohlfarth astroid. The laser pulses were obtained from a femtosecond laser (500 fs pulse width, 300 kHz repetition rate, 1040 nm center wavelength, ~15 nJ pulse energy) and focused to a $1/e^2$ diameter of ~8 µm on the top Al layer, see Fig. 1(c). A more detailed description of the sample properties can found in a previous study.[20] Although the measurements detailed in this paper have been conducted on the nanopillar stack described above, samples with different nanopillar dimensions showed similar results.

To obtain information about the acoustic pulses generated in the Al layer by the femtosecond laser pulses, we performed a pump and probe reflectometry experiment.[21] Therefore, the pump laser pulse was focused on the Al layer generating acoustic pulses via thermoelastic coupling. The probe pulse, which can be time delayed with respect to the pump pulse, was focused on the Al layer next to the pump beam and the reflection change of the probe beam was recorded versus time delay between pump and probe beams. In principle, both, picosecond strain pulses (propagating into the sample) and SAWs (propagating along the surface) can be detected. We will show below that for the excitation of magnetization of precession only SAWs are relevant. The time traces of SAWs for three different distances (5.3 µm, 8.9 µm, and 13.4 µm) between pump and probe pulses are shown in Fig. 1(b). The SAWs have a bipolar shape which is similar to the shape of a typical picosecond strain pulse propagating into the sample but with much longer duration of several ns. The duration difference between the picosecond strain pulses and the nanosecond SAW is related to the thermal distribution in the Al layer after laser excitation (which extends several µm along the surface but only several 10 nm normal to the surface). Comparing the time delay between the measured SAWs with the distances between pump and probe, we estimate the SAW velocity to be (3.3±0.5) µm/ns. This value compares well with literature data for SAWs in Al of 2.95 µm/ns.[22]

The tunnel resistance of MTJs depends on the angle $\phi$ between the magnetization orientation of the free layer and the fixed layer[23]:

$$R(\phi) = R_{\perp}[1 + B\cos(\phi)]^{-1}, \quad (1)$$

Where $B = \frac{R_{\mathrm{AP}} - R_{\mathrm{P}}}{R_{\mathrm{AP}} + R_{\mathrm{P}}}$, $R_{\perp} = \frac{2R_{\mathrm{AP}}R_{\mathrm{P}}}{R_{\mathrm{AP}} + R_{\mathrm{P}}}$ and $R_{\mathrm{P}}$ and $R_{\mathrm{AP}}$ are the resistance values for $\phi = 0°$ (parallel alignment, P) and $\phi = 180°$ (antiparallel alignment, AP), respectively. Due to the magnetoelastic effect, a phonon pulse leads to an angular excursion of magnetization of the free layer, which in turn causes a change of the tunnel resistance. We neglect the magnetoelastic effect in the fixed layer, since its magnetization is well pinned by the synthetic antiferromagnet. The measurement of SAW-induced magnetization dynamics in the MTJ is realized by time resolved measurements of tunnel resistance changes. It is worth to mention that the penetration depth of SAW into the substrate is close to its wavelength $\lambda \approx$ several µm.[24] The MTJ is located just 150 nm below the top Al layer and, thus, well within the penetration depth of the SAWs.

To study the SAW induced magnetization dynamics in the MTJ, we measured the time resolved voltage change under constant bias current due to tunnel resistance changes by using a sampling oscilloscope with 50 Ω input impedance. The trigger signal for the oscilloscope was obtained from the femtosecond laser system such that the oscilloscope time axis is synchronized with the laser pulses. A small current ($I_{DC}$ = ±400 µA) was applied to the MTJ through a bias tee,[16] see Fig. 1(a). To separate signals due to tunnel resistance changes from unwanted background signals, two oscilloscope traces taken for $+I_{DC}$ and $-I_{DC}$ were subtracted from each other after averaging over 2000 individual traces. Static magnetic fields up to 300 Oe at various in-plane angles θ were applied as indicated in Fig. 1(c).

We now comment on the experimental results. Figures 2(a) and (b) show the measured oscilloscope traces for two different magnetic field amplitudes and two different laser excitation positions. While the magnetic field was either $H_{\theta=85^\circ}$ = 120 Oe (red, lower curves), corresponding to an AP state of the MTJ, or $H_{\theta=85^\circ}$ = -150 Oe (black, upper curves), corresponding to a P state of the MTJ, the excitation spots were either right above the MTJ (a) or 10 µm away (b). In all cases, an oscillatory behavior due to precession is observed. The magnetization precession is approximately a factor of three larger in the AP state than in the P state and it maintains several nanoseconds due to the long SAW duration, see Fig. 1(b). The difference between the AP and P state is linked to the dependence on the applied magnetic field angle, which will be discussed below. Comparing the precession obtained for the different laser excitation spots it is obvious that for excitation 10 µm away from the MTJ nanopillar, the precession starts about 3 ns later as compared to an excitation right above the MTJ. This time delay agrees very well with the propagation time of the SAWs, see Fig. 1(b), underlining that it is indeed the SAW which induces the magnetization dynamics. We can exclude the strain pulse propagating into the substrate as being responsible for the magnetization precession, since it only takes about 80 ps for the strain pulse to propagate from the top Al layer to the MTJ.

For laser excitation right above the MTJ, the AP signal experiences a slow decay within the first ns after the laser pulse hits the Al contact, see Fig. 2(a). This is not observed in Fig. 2(b) where the red curve is shifted along the y axis for clarity. The slowly decaying signal in Fig. 2(a) is due to the heat diffusion from the Al surface to the buried MTJ and results from temperature dependence of spin polarized tunneling. Using a previously published method,[20] we estimate the time-dependent temperature rise in the MTJ to be approximately 3.2 K. We checked the temperature dependence of the precession frequency obtained from STT experiments with step-like voltage pulses using an electric heating stage below the MTJ sample. A temperature increase up to 30 K has almost no influence on the precession. Due to this dependence and because the slowly decaying contribution of Fig. 2(a) vanishes for an excitation position ~10 µm away from the MTJ, we can safely exclude the laser-induced temperature rise and, thus, a thermal STT as the origin of the observed magnetization dynamics.

To further study the SAW induced magnetization dynamics, the applied magnetic field amplitude and angle was systematically varied for laser excitation right above the MTJ. We find that the largest precession amplitude occurs for a magnetic field of approximately 150 Oe (AP state), see Fig. 2(c). The magnetic field dependence is attributed to a resonance between the induced magnetoelastic mode and the SAW frequency.[12,25,26] In addition to the dependence on magnetic field amplitude we also find a pronounced dependence on the magnetic field angle, see Fig. 2(d). The largest precession occurs for an applied field close to the hard axis of the MTJ ($\theta = 85°$). The amplitude of precession gradually decreases when the magnetic field is changed from $\theta = 85°$ to $\theta = 0°$. At $\theta = 0°$, only small magnetization precession can be found in a small magnetic field range which is close to the switching field. We believe that this angle dependence of the precession signal mainly indicates the existence of non-uniform modes localized close to the edges of a ferromagnetic stripe. These modes typically occur for external magnetic fields applied perpendicular to an anisotropy field.[17,27] We will comment on the existence of non-uniform modes in more detail further below.

We now analyze the dependence of the precession frequency on the SAW stimulus. This analysis is important to identify the type of magnetization modes being excited by the SAW. In Fig. 3(a) we have plotted the precession frequency of the magnetoelastic mode versus applied magnetic field amplitude for $\theta = 85°$ and two optical excitation energies (blue squares and red triangles for 7.5 nJ and 15 nJ, respectively). The right-hand-side inset of Fig. 3(a) shows the precession frequency spectrum versus magnetic field for 15 nJ excitation pulse energy. In general, the precession frequency increases with applied field amplitude and does not depend on the optical excitation power. The latter dependence further demonstrates that the SAW induced precession is not the result of optical heating of the sample. In such a case we would expect a pronounced dependence on optical excitation power.

In previous studies on magnetoeleastic effects in thin magnetic layers, precession signals were read out using optical techniques[6,11,14,25,28] and it was difficult to determine the spin wave mode of the unperturbed and perturbed system (with respect to elastic perturbations). In our work we measure the precession using fast electrical read out of the MTJ resistance. Since its magnetization dynamics driven by charge current pulses has been well studied[16,18] our experimental scheme allows for the comparison of the unperturbed magnetization modes with the SAW induced modes. With this comparison we are able to assign the SAW-induced magnetization mode to a spatially non-uniform spin wave mode being mainly localized at the edges of the free layer as explained in the following.

We applied 180-ps-long current pulses with an amplitude of approximately 8 mA and a repetition rate of 100 kHz to the MTJ. The free magnetization precession induced by the current pulse through STT was measured after the pulse decay using a fast sampling oscilloscope[16,20] with the magnetic field applied along the hard axis. The precession frequencies obtained from this experiment are visualized in Fig. 3(a) as black circles versus magnetic field amplitude. The left-hand-side inset of Fig. 3(a) shows the precession frequency spectrum versus magnetic field. While the magnetic field dependence of the free precession qualitatively resembles the magnetic field dependence of the SAW induced precession, the SAW frequencies are always larger than the free precession frequencies. Most likely, this frequency difference results from the transition of pure spin waves to magnetoelastic waves. Calculating the dispersion relation of magnetoelastic waves one finds that in the crossover region between pure spin waves and pure elastic waves two magnetoelastic branches exist, having larger and smaller frequencies than the pure spin wave.[29] We believe that the SAW mainly excites the larger frequency branch of the magnetoelastic mode. We can rule out that the magnetization change induced by the magnetoelastic effect causes a significant shift of the precession frequency of a certain magnetization mode. If this were true, we would have observed a dependence of the precession frequency on the optical excitation energy in Fig. 3(a).

Since the magnetic field dependence of the SAW induced precession closely resembles the free precession, it is very likely that the magnetization modes are equal. It is therefore possible to simulate the free precession in the free layer to obtain qualitative information about the magnetization mode being induced by the SAW. A detailed calculation of the magnetoelastic mode is beyond the scope of this paper. For the simulation of free precession we have employed the micromagnetic simulation tool mumax[30]. The following parameters have been used for the simulation: saturation magnetisation $M_s$ = 796 kA/m, perpendicular magnetic anisotropy $K_u$ = 7.96 kJ/m$^3$, and exchange constant $A_{ex}$ = 20 pJ/m. The simulated sample, which has the same nominal dimensions as the in the experiment, is discretized in elements of 2 nm × 2 nm × 2.6 nm. Figure 3(b) shows the simulated free precession frequency versus applied magnetic field along the y direction for an excitation of the free layer with a sinc function having a cut-off frequency of 15 GHz and a total simulation time of 50 ns. The simulated behavior qualitatively agrees with the measurements. However, the simulated frequencies are higher than experimentally observed. We attribute this difference to certain parameters of the simulation which are not exactly known such as the exact shape of the free layer (e.g., deviation from the nominal shape after lithography). The upper inset of Fig. 3(b) shows the y component of the static magnetization vector $\mathbf{m}_{eq}(x,y)$ of the magnetization mode at an applied magnetic field of 225 Oe. The lower inset shows the y component of the dynamical part $\boldsymbol{\delta}\mathbf{m}(x,y)$ of the magnetization mode, which is obtained from the magnetization $\mathbf{m}$ at a certain time instant using $\boldsymbol{\delta}\mathbf{m} = \mathbf{m} - (\mathbf{m}\cdot\mathbf{m}_{eq})\mathbf{m}_{eq}$. Both, the static and the dynamical parts clearly show that the free precession mode is confined close to the edges of the free layer. This, in turn, strongly suggests that also the SAW induced magnetization dynamics in the free layer of the MTJ is linked to a spatially non-uniform mode being localized close to the edges.

Finally, the coherent nature of SAW-induced magnetization dynamics also enables the coherent manipulation of the magnetization by means of two SAW pulses. Figure 4(a) shows time-resolved magnetization traces (again obtained from tunnel resistance measurements using a fast sampling oscilloscope) employing two laser pulses, which were focused onto the same position 10 µm away from MTJ and can be time delayed with respect to each other. A clear periodic dependence of destructive and constructive interferences of magnetic oscillations on the delay between two laser pulses is observed, revealing that we can either amplify or quench the precession by coherent control. It should be noted that coherent control of the magnetization precession can also be achieved when keeping the time delay between the two laser pulses constant and varying the position of one laser spot with respect to the other. The measured interference pattern agrees very well with a calculated superposition of two separately measured magnetization traces; see Fig. 4(b). The coherent control study directly extends previous coherent control experiments on magnetization dynamics[2,4,31–33] to an industrially relevant device and, thus, might prove useful for future application.

In summary, we have employed MTJs to study magnetization dynamics driven by femtosecond-laser-pulse-induced SAWs. We could identify a spatially non-uniform magnetization mode as being excited by the SAWs and demonstrated coherent control of magnetization in MTJs using acoustic pulses. Our results open prospects for future applications, in which magnetization has to be controlled on ultrafast time scales. Additionally, they provide valuable information for spincaloritronic studies in which temperature and temperature gradients are generated by excitation with ultrafast optical pulses. Our time-resolved experiments directly show that the optically generated acoustic pulses must not be neglected and, under certain experimental conditions, even fully determine the optically induced magnetization dynamics.

The authors thank Piet Schmidt for the loan of the laser system used in the present work and Erik Benkler for technical assistance. Funding by the European Metrology Research Programme (EMRP, Joint Research Project EXL04, RMG 15SIB06-RMG2) is gratefully acknowledged. The EMRP is jointly funded by the EMRP participating countries within EURAMET and the European Union. J.D.C. is thankful for the support of FCT grant SFRH/BD/7939/2011.

**Figure 1**

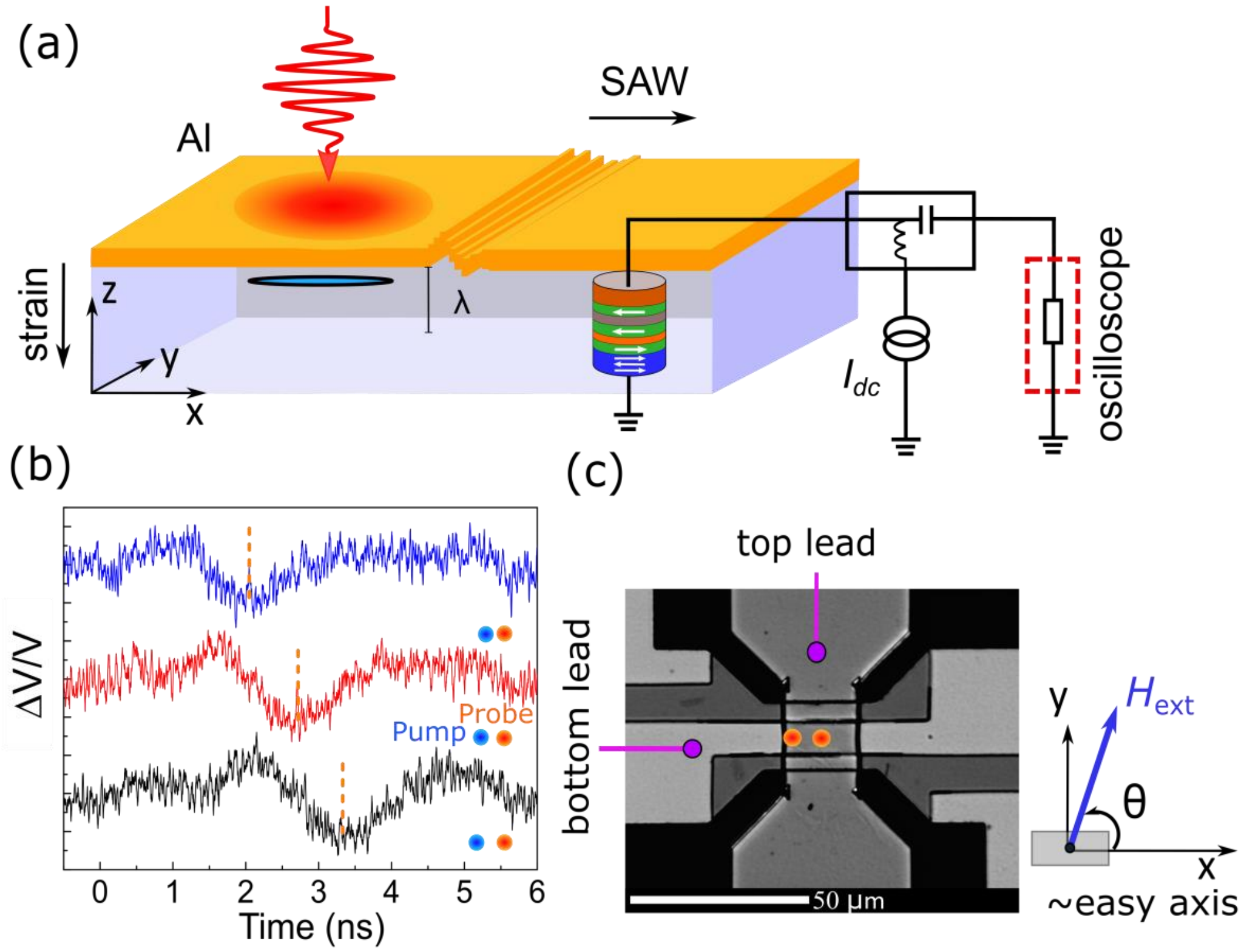

Fig. 1. (a) Schematic sketch of the experimental setup. (b) Time-domain reflectance measurements of SAWs induced by femtosecond laser pulses for three different distances (5.3 µm, 8.9 µm, and 13.4 µm) between pump and probe pulse. The dashed vertical lines denote the minima of the SAWs used to calculate the SAW velocity. (c) Microscope image of the electrical contacts above the MTJ nanopillar with two different laser heating positions (orange dots) and orientation of the MTJ's easy axis as well as the externally applied in-plane magnetic field.

**Figure 2**

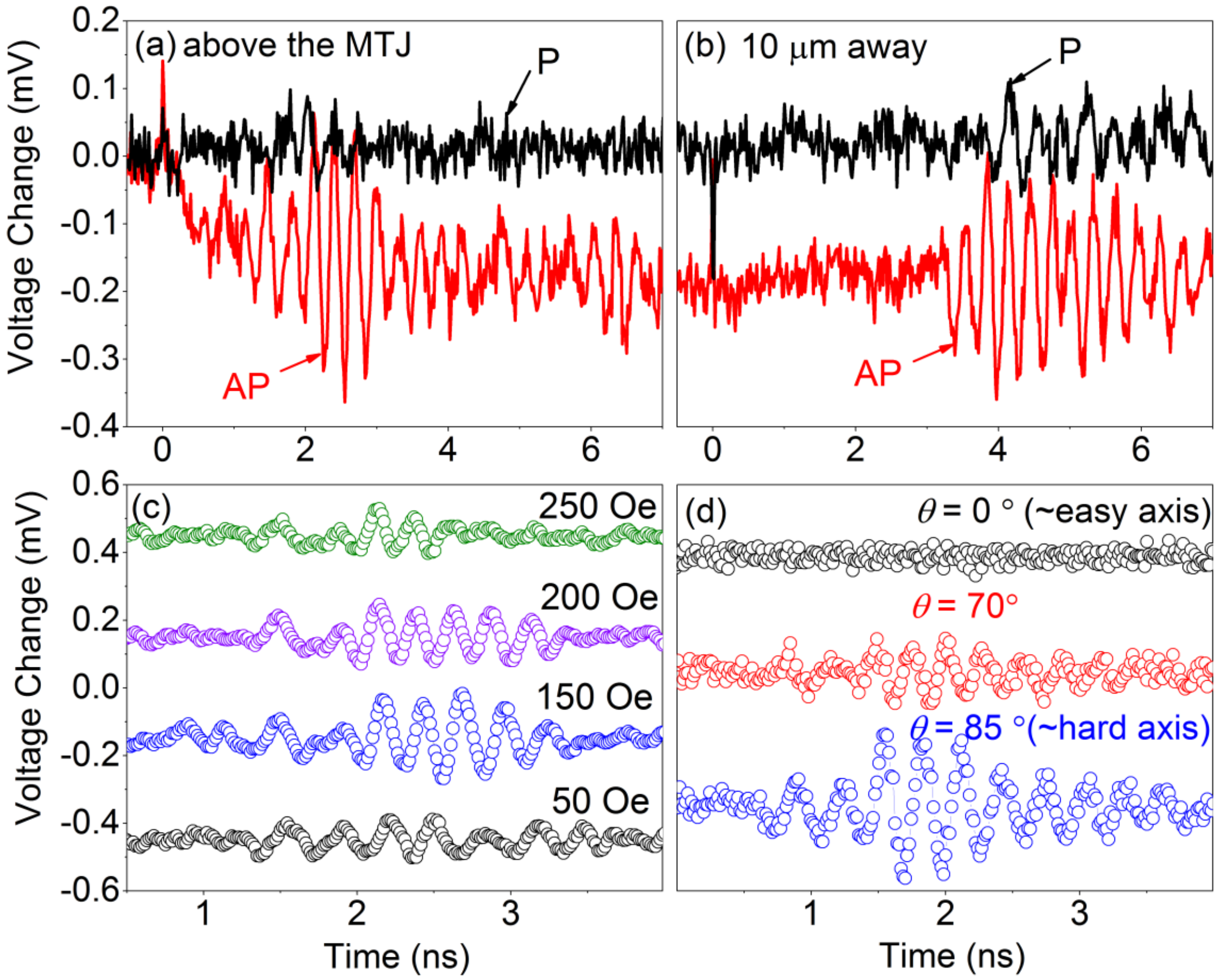

Fig. 2. Time-resolved voltage change across the MTJ for different excitation conditions and magnetic fields. (a) Optical excitation right above the MTJ nanopillar and (b) 10 µm away from the nanopillar. The black and red traces correspond to parallel (P, $H_{ext}$=-150 Oe) and antiparallel (AP, $H_{ext}$=120 Oe) alignment of the MTJ for θ = 85°. In (b) the red curve is shifted by -0.2 mV for clarity. (c) Optical excitation right above the MTJ and different magnetic field amplitudes for θ = 85°. All curves are shifted along the y axis for clarity. (d) Optical excitation right above the MTJ and a magnetic field amplitude of $H_{ext}$=120 Oe but for different field angles θ. The black and blue curves are shifted along the y axis for clarity.

**Figure 3**

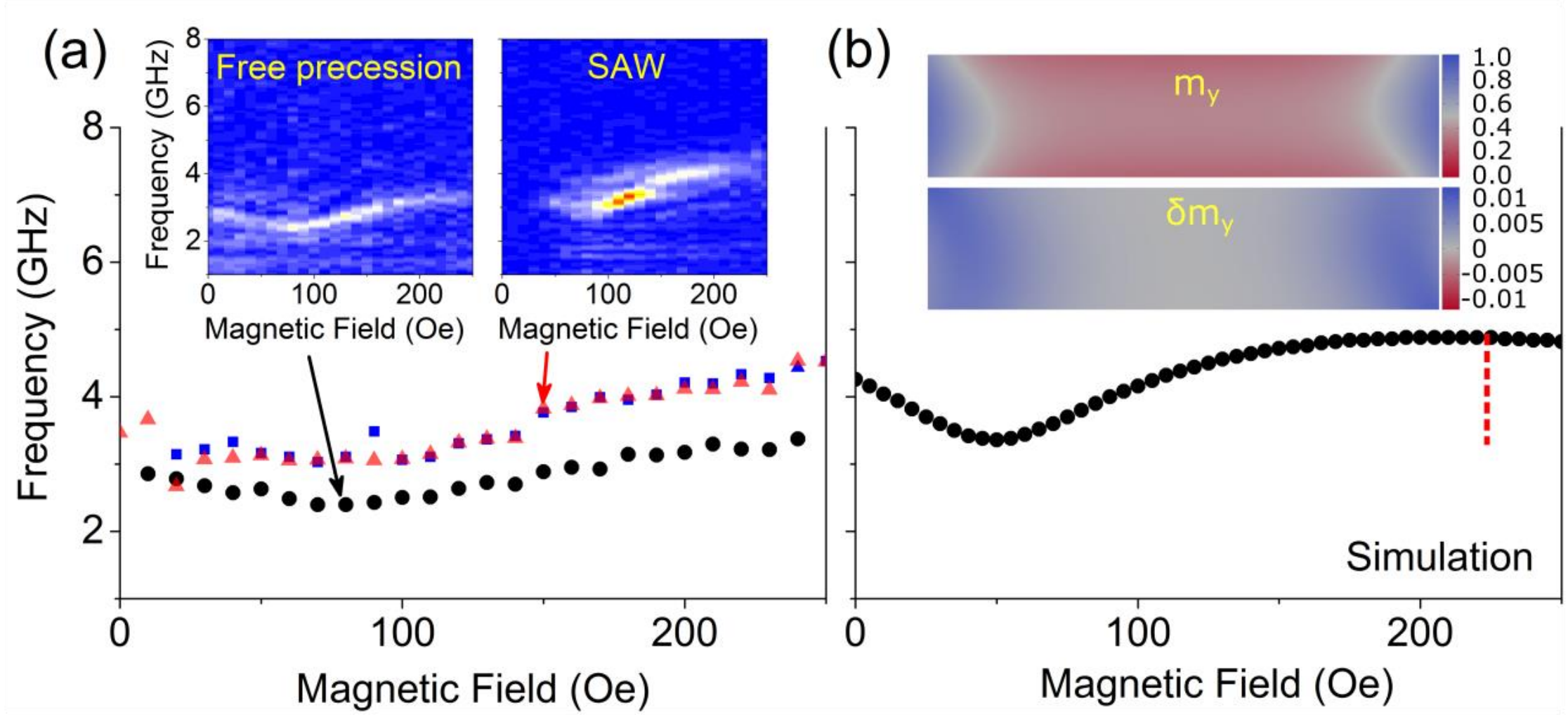

Fig. 3. (a) Precession frequency versus applied magnetic field amplitude along the hard axis for two different optical excitation energies (blue squares for 7.5 nJ and red triangles for 15 nJ) and for excitation with 180-ps-long current pulses (black dots). The insets show the corresponding frequency spectra versus magnetic field. (b) Simulated frequency of the free precession in the free layer versus applied magnetic field amplitude. Upper inset: Simulated equilibrium configuration of the magnetization component in the y direction in the free layer for an applied magnetic field amplitude of 225 Oe. Lower inset: Dynamical changes of the magnetization component in the y direction in the free layer for an applied magnetic field amplitude of 225 Oe.

**Figure 4**

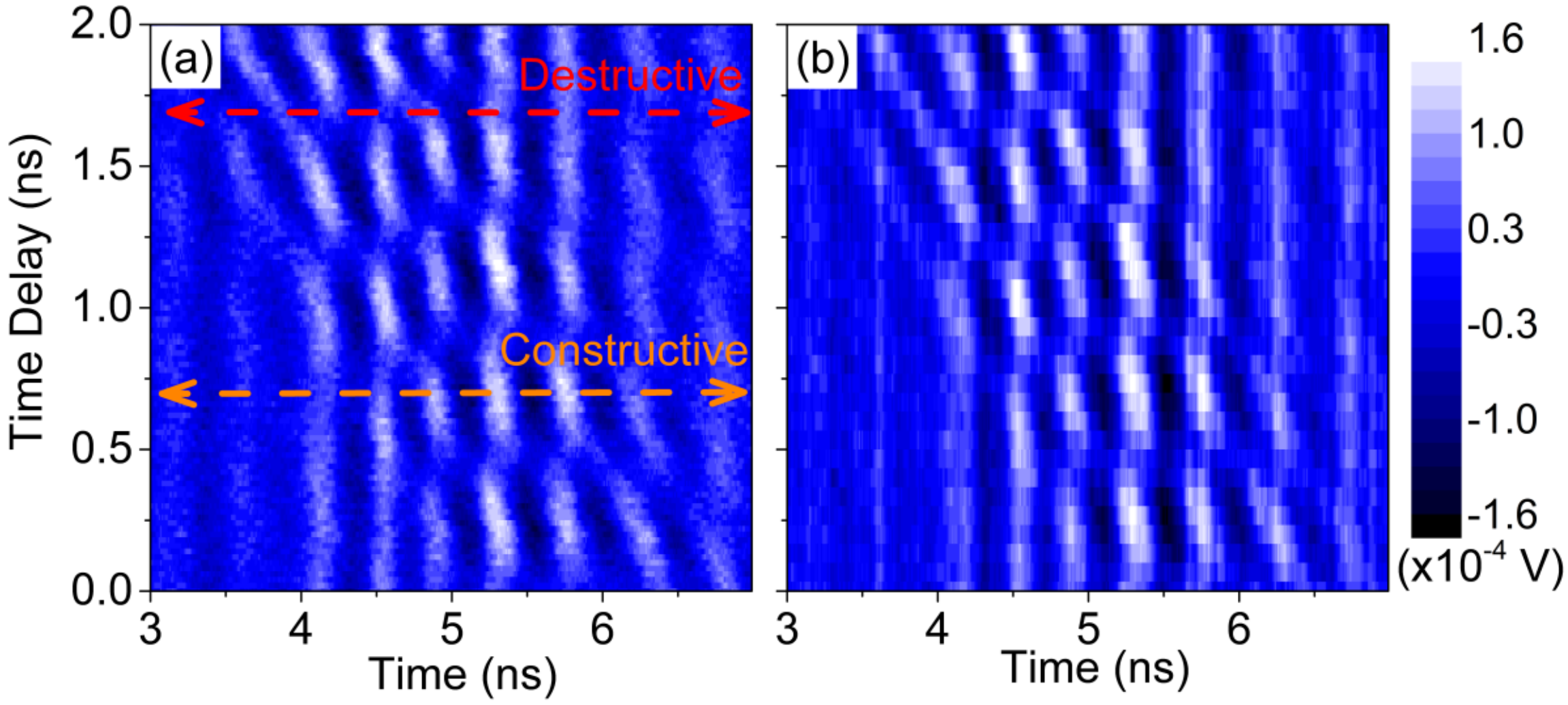

Fig. 4. (a) Coherent control of magnetization dynamics using two time-delayed laser pulses. (b) Calculated magnetization dynamics from superposition of two separately measured magnetization traces, where the time delay of one laser pulse is kept fix and the other is subsequently shifted in time.